\documentclass[a4paper,12pt]{article}
\usepackage{graphicx}
\usepackage{float}
\begin{document}

\date{}

\title{\bfseries{Resonant Coupling in the Heteronuclear Alkali Dimers for Direct Photoassociative Formation of X(0,0) Ultracold Molecules}}

\author{W.C. Stwalley, J. RayMajumder, M. Bellos, R. Carollo, \\M. Recore, M. Mastroianni\\ \\{\itshape{University of Connecticut, Storrs, CT 06269, USA}}\\ \\ {\bfseries E-mail}: w.stwalley@uconn.edu}

\maketitle

\begin{abstract}

\thispagestyle{empty}

Promising pathways for photoassociative formation of ultracold heteronuclear alkali metal dimers in their lowest rovibronic levels (denoted X(0,0)) are examined using high quality {\itshape {ab initio}} calculations of potential energy curves currently available.  A promising pathway for KRb, involving the resonant coupling of the $2 ^1\Pi$  and $1 ^1\Pi $  states just below the lowest excited asymptote (K($4s$)+Rb($5p_{1/2}$)), is found to occur also for RbCs and less promisingly for KCs as well.  The resonant coupling of the $3 ^1 \Sigma ^+ $ and $1 ^1\Pi $  states, also just below the lowest excited asymptote, is found to be promising for LiNa, LiK, LiRb, and less promising for LiCs and KCs.  Direct photoassociation to the $1 ^1\Pi $ state near dissociation appears promising in the final dimers, NaK, NaRb, and NaCs, although detuning more than 100 cm$^{-1}$ below the lowest excited asymptote may be required.
\end{abstract}

\pagebreak

\section{Introduction}

Research with ultracold atoms has been an amazingly fruitful area of frontier research for over two decades, with many exciting developments still occuring frequently.  Ultracold research has expanded into other areas of atomic, molecular, and optical physics including ultracold ions, plasmas, and especially molecules, where similarly exciting developments are now occuring \cite{Krems}.

In ultracold molecules, emphasis has shifted to ultracold polar molecules, e.g. heteronuclear alkali metal dimers, because of the permanent dipole moments of these molecules, which vary from small (KRb) to large (LiCs) \cite{Aymar}. Emphasis has also shifted from study of the weakly bound levels formed by photoassociation (PA) (1-100 cm$ ^{-1}$ below dissociation asymptotes) and extremely weakly bound levels formed by magnetoassociation (MA) via Feshbach resonances to formation of the most strongly bound levels, i.e. the rovibronic ground state, X$ ^1\Sigma ^+ $ ($v$=0, $J$=0), or X(0,0) for short. [Actually this lowest $``$level$"$ consists of a large number of hyperfine states with very small splittings not normally observable in high resolution spectroscopy \cite{Bahns}.]  The major motivation for seeking X(0,0) molecules is their lack of inelastic collisions at ultracold temperatures [although the hyperfine state distribution can certainly be redistributed by inelastic collisions].  A second motivation is that the dipole moment is expected to be largest for the X(0,0) level, an order of magnitude larger than for levels formed by PA and many orders of magnitude larger than for levels formed by MA \cite{Zemke}.

The formation of X(0,0) levels in ultracold molecules was first achieved in K$_2$ by spontaneous emission following two-photon PA \cite{Nikolav} and then in RbCs using stimulated Raman following spontaneous emission following PA \cite{Sage} and in LiCs by spontaneous emssion following far-detuned PA \cite{Deiglmayr}.  More recently, KRb formed by MA has been efficiently transferred by stimulated Raman to X(0,0) \cite{Ni} and Cs$_2$ formed by PA has been optically pumped to X(0,0) \cite{Viteau}.  Further exciting results are expected in the very near future.

Two alternate approaches for forming X(0,0) molecules have recently been proposed:  Feshbach Optimized Photoassociation (FOPA) \cite{Pellegrini} and Resonant Coupling of PA Excited States \cite{Dion}-\cite{Wang}.  It is the latter approach which we will explore here for all 10 heteronuclear alkali metal dimers,  beginning with the best understood case of KRb, which we have studied extensively at UConn \cite{Wang}.  It should be noted that FOPA could be described as $``$Resonant Coupling of PA Lower States$"$ and is closely related to our approach.

Resonant coupling of levels in two (or more) electronic states is, of course, a very well known and long explored topic in electronic spectroscopy;  we say the levels mutually perturb each other or that the levels are of mixed character.  When levels from different electronic states have nearly the same energy and quantum numbers, their wavefunctions become mixtures of single electronic channel wavefunctions.  A particularly well studied example is the resonant coupling of the levels of the A  $^1\Sigma_u ^+$  and  b$ ^3\Pi_u$ states of Na$_2$ \cite{Qi}.  This coupling has been known since the work of R. W. Wood over a century ago.  However, the resonant couplings we are looking for are those which: (1) occur in the PA energy regions ($<$100 cm $^{-1}$ below the PA asymptote); and (2) have good Franck-Condon wavefunction overlap with the X(0,0) wavefunction. The next section describes the very promising example of KRb and then we shall explore the other nine heteronuclear dimers.   

\section{The Example of KRb}

\begin{figure}[h]
\begin{center}
\includegraphics[width=4.0 in]{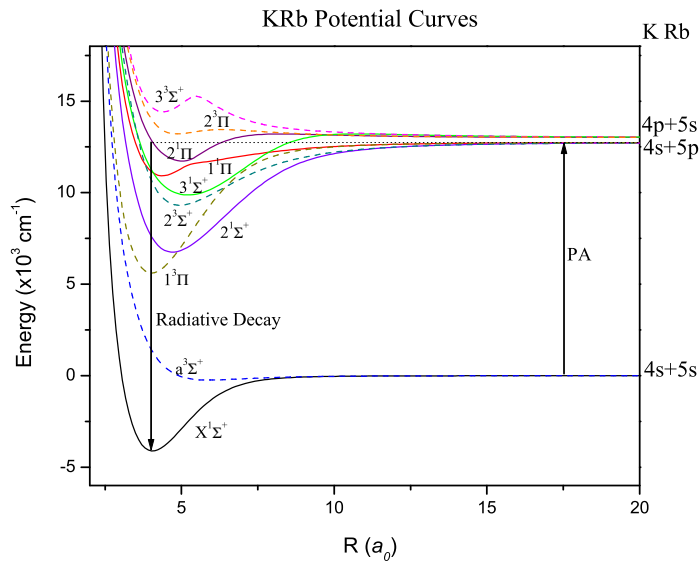}
\caption{Potential energy curves for KRb up to the second exited atomic asymptote based on the data in Ref.\cite{Rousseau}.  The vertical arrows correspond to Photoassociation (PA) and Radiative decay.  The horizontal dotted lines correspond to the energy of the lowest excited atomic asymptote.}
\label{graph:KRb}
\end{center}
\end{figure}

As discussed more completely elsewhere \cite{Wang}, the potential energy curves of KRb are fairly well known from several sets of high-quality ab initio calculations (e.g.  Ref.\cite{Rousseau}) and from experiment (e.g. Refs. \cite{Okada} and \cite{Kasahara}) . A set of {\itshape {ab initio}} curves including spin-orbit interactions \cite{Rousseau}  is shown in Fig.\ref{graph:KRb}.

\begin{figure}[h]
\begin{center}
\includegraphics[width=4 in, height= 3 in]{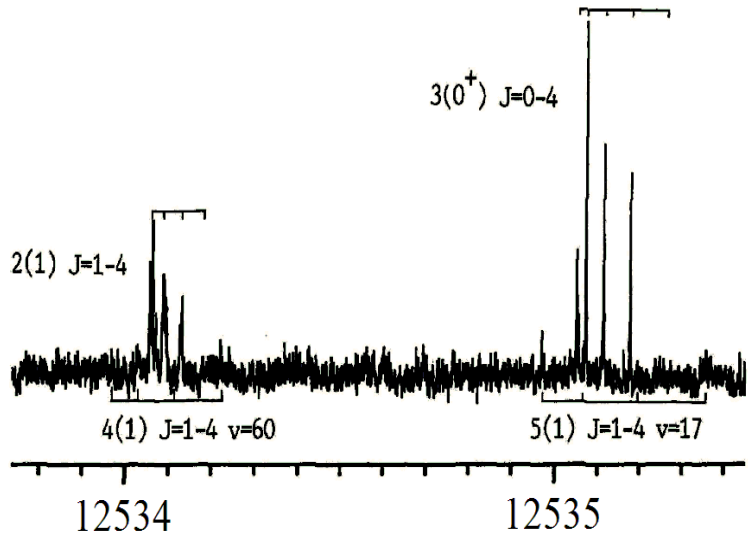}
\caption{Fragment of the PA spectrum of $^{39}$K $^{85}$Rb from Ref.\cite{Wang} in the 12533.8-12535.4 cm$^{-1}$ range, showing two previously assigned bands \cite{Wang} of the 2(1) and 3(0$^+$) states (marked above) and two newly assigned bands of the 4(1)[$1 ^1\Pi$] and 5(1)[$2 ^1\Pi$] states (marked below), corresponding to resonantly coupled vibrational levels v=60 and v=17, respectively.}
\label{graph:spectrum}
\end{center}
\end{figure}

\begin{figure}
\begin{center}
\includegraphics[width= 5 in, height= 6 in]{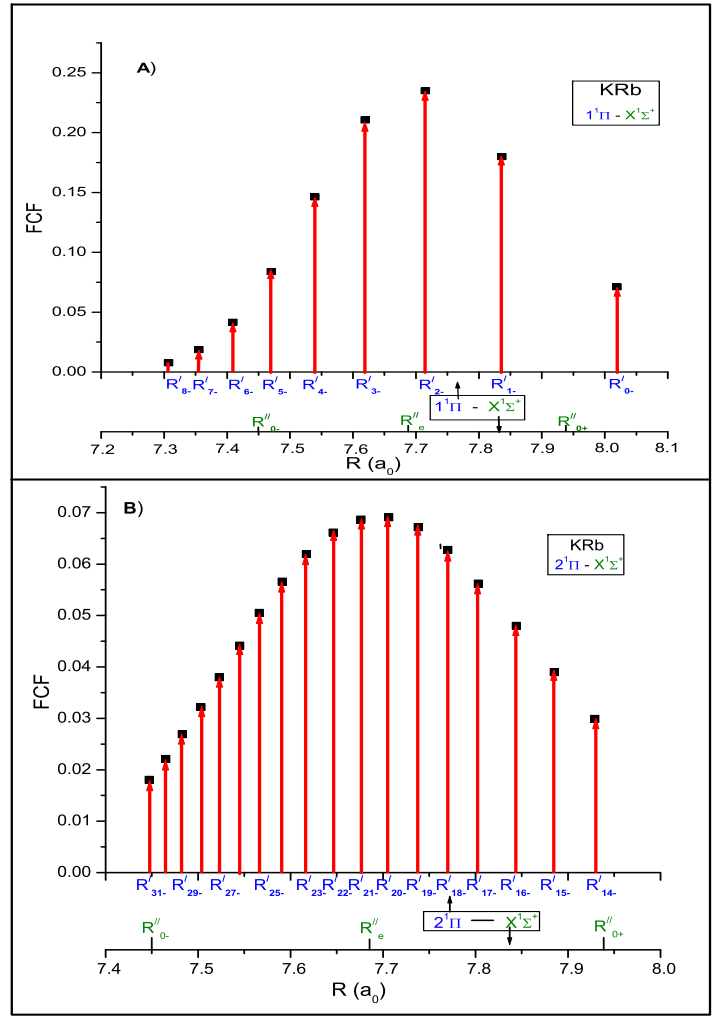}
\caption{FCFs versus internuclear distances for A) $1 ^1\Pi$ ($v'$) - $X ^1\Sigma^+$ ($v''$=0) transitions and B) $2 ^1\Pi$ ($v'$) - $X ^1\Sigma^+$ ($v''$=0) transitions. . The blue coordinates are the inner classical turning points $R_{v-}'$ of the upper state and the green are the inner turning point $R_{0-}''$, the equilibrium internuclear distance $R_{e}''$, and the outer turning point $R_{0+}''$ of the X(0,0) level. All internuclear distances are represented on the same scale, shown at the bottom.}
\label{graph:CondonReflection}
\end{center}
\end{figure}

Kato and co-workers extensively studied the $1 ^1\Pi$ and $2 ^1\Pi$ states and their mutual perturbations \cite{Okada},\cite{Kasahara}.  Their studies included levels of both states in the PA energy region below the K($4s$) + Rb($5p_{1/2}$) asymptote and above it; those above the asymptote were observed to predissociate. The $1 ^1\Pi$ state correlates with the higher K($4s$)+ Rb($5p_{3/2}$) asymptote while the $2 ^1\Pi$ state correlates with the much higher K($4p_{3/2}$)+ Rb($5s$) asymptote.  The PA energy region levels of both states were tabulated in Ref.\cite{Wang}, and the coupled levels noted.  Our initial assignments of our KRb   PA spectra \cite{Wang} included observations of v=61-63 of the $1 ^1\Pi$ state, but these levels are not resonantly coupled to levels of the $2 ^1\Pi$ state.  Recently we have identified levels v=60 of the $1 ^1\Pi$ state and v=17 of the $2 ^1\Pi$ state, which are resonantly coupled, as shown in Fig.\ref{graph:spectrum}.  It is the wavefunction of the v=17 level which has good Franck-Condon overlap with the wavefunction of the X(0,0) level.  However, without resonant coupling we would not have access to this level by PA.  That is because the v=17 level is basically a short range level with an extremely small probability of being reached by PA, since the $2 ^1\Pi$ potential curve is dissociating to a much higher asymptote and the classically forbidden part of the v=17 wavefunction starts at $\sim$11 $a_0$ (see Fig.\ref{graph:CondonReflection} of Ref.\cite{Wang}).  However, since v=17 and v=60 are strongly mixed, we expect the former to provide good overlap with X(0,0) and the latter to provide good PA probability.

In the following section, we examine high quality {\itshape {ab initio}} potential energy curves for the nine other heteronuclear alkali metal dimers \cite{Petsalakis}-\cite{Allouche} and discuss their promise for similar resonant-coupling-assisted formation of X(0,0) dimers. This is in part due to the paucity of high quality spectroscopic results for these species in the PA energy region similar to those for KRb \cite{Okada},\cite{Kasahara}. We also note that calculations including spin-orbit coupling are not generally available and thus the arguments are somewhat less compelling for the species containing heavier alkali atoms, especially Cs.

We note that it is more efficient to estimate strong transitions by simply examining the internuclear distances at which classical inner and outer turning points occur rather than generating tables of Franck- Condon factors. Thus, for example, the photoassociation of $Cl_2$  was estimated by examining the upper state wavefunctions with good overlap with the $v''$ wavefunctions \cite{Gibson}, as discussed in Ref.\cite{Herzberg}( especially in connection with Fig. 175).

Thus to strengthen our arguments, we have plotted the Franck-Condon factors (FCFs) for $1 ^1\Pi$ ($v'$) - $X ^1\Sigma^+$ ($v''$=0) transitions Fig.(\ref{graph:CondonReflection}A) and $2 ^1\Pi$ ($v'$) - $X ^1\Sigma^+$ ($v''$=0) transitions Fig.(\ref{graph:CondonReflection}B). As can be clearly seen in Fig.\ref{graph:CondonReflection}A the peak of the FCFs for $1 ^1\Pi$ ($v'$) - $X ^1\Sigma^+$ ($v''$=0) transitions is at $v'$=2, which would be very difficult to reach by direct PA. On the other hand, for $2 ^1\Pi$ ($v'$) - $X ^1\Sigma^+$ ($v''$=0) transitions, the peak is at $v'$=20 and $v'$=17 also has a very promising Franck-Condon overlap with X(0,0) level. However, because of the strong mixing (resonant coupling) of $v'$=17 level of the $2 ^1\Pi$ state with $v'$=60 level of the $1 ^1\Pi$ state and the observation of these two mixed levels in Fig.\ref{graph:spectrum}, these mixed levels will each have good FCFs for emission to the X(0,0) level.  
 
\section{Other Heteronuclear Dimers}

\subsection{  LiNa\cite{Petsalakis}}

\begin{figure}[h]
\begin{center}
\includegraphics[width=4 in, height=3 in]{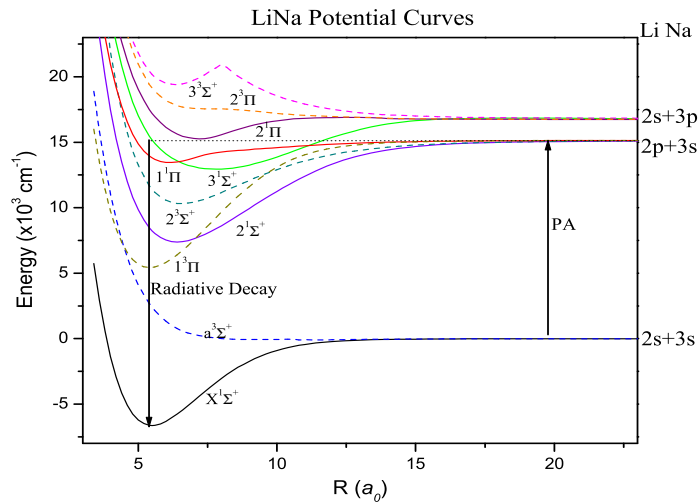}
\caption{Potential energy curves for LiNa up to the second exited atomic asymptote based on the data in Ref.\cite{Petsalakis}.  The vertical arrows correspond to Photoassociation (PA) and Radiative decay.  The horizontal dotted lines correspond to the energy of the lowest excited atomic asymptote.}
\label{graph:LiNa}
\end{center}
\end{figure}

The potential energy curve which is most promising for emission to X(0,0) is the $3 ^1\Sigma ^+$ state (Fig.\ref{graph:LiNa}).  However, since direct PA to this short range state would be expected to be very weak, it is only with resonant coupling with the $1 ^1\Pi$ long range state that this becomes a promising pathway.  Note that the two states cross each other twice, at approximately 6.5 and 11 $a_o$.

Note that there is still an avoided crossing between the $2 ^1\Pi$ and $1 ^1\Pi$ states at $\sim$7.5 $a_o$.  However, the resonantly coupled levels will be above the lowest excited asymptotes Li($2p_J$)+Na($3s$).  The possibility of direct PA to the $2 ^1\Pi$ state just below the Li($2s$)+Na($3p_J$) asymptotes is not as promising as it might appear because the state is repulsive\cite{Kasahara} at large internuclear distance;  PA through a quasibound resonance might however be possible.

\subsection{  LiK[Previously unpublished calculations by Olivier Dulieu]}

\begin{figure}[h]
\begin{center}
\includegraphics[width=4 in, height=3 in]{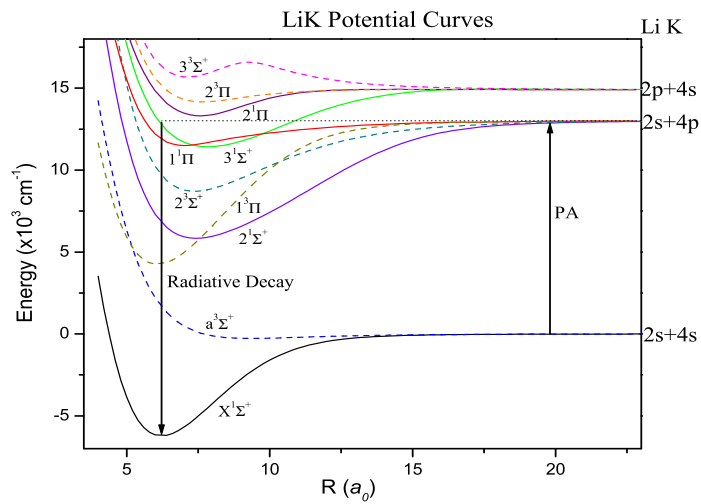}
\caption{Potential Energy curves for LiK up to the second exited atomic asymptote based on the unpublished calculations by Olivier Dulieu.  The vertical arrows correspond to Photoassociation (PA) and Radiative decay.  The horizontal dotted lines correspond to the energy of the lowest excited atomic asymptote.}
\label{graph:LiK}
\end{center}
\end{figure}

The potential energy curve which is most promising for emission to X(0,0) is again the $3 ^1\Sigma^+$ state (Fig.\ref{graph:LiK}).  Again, direct PA to this short range state would be expected to be extremely weak.  Also resonant coupling with the $1 ^1\Pi$ long range state makes this a promising pathway. The two states cross twice at $\sim$7 and 10 $a_o$. 

Again there is an avoided crossing of the $2 ^1\Pi$ and $1 ^1\Pi$ states, but the coupled levels are above the lowest excited asymptotes Li($2s$)+K(4$p_{J}$).

\subsection{  LiRb\cite{Korek001}}

\begin{figure}[h]
\begin{center}
\includegraphics[width=4 in, height=3 in]{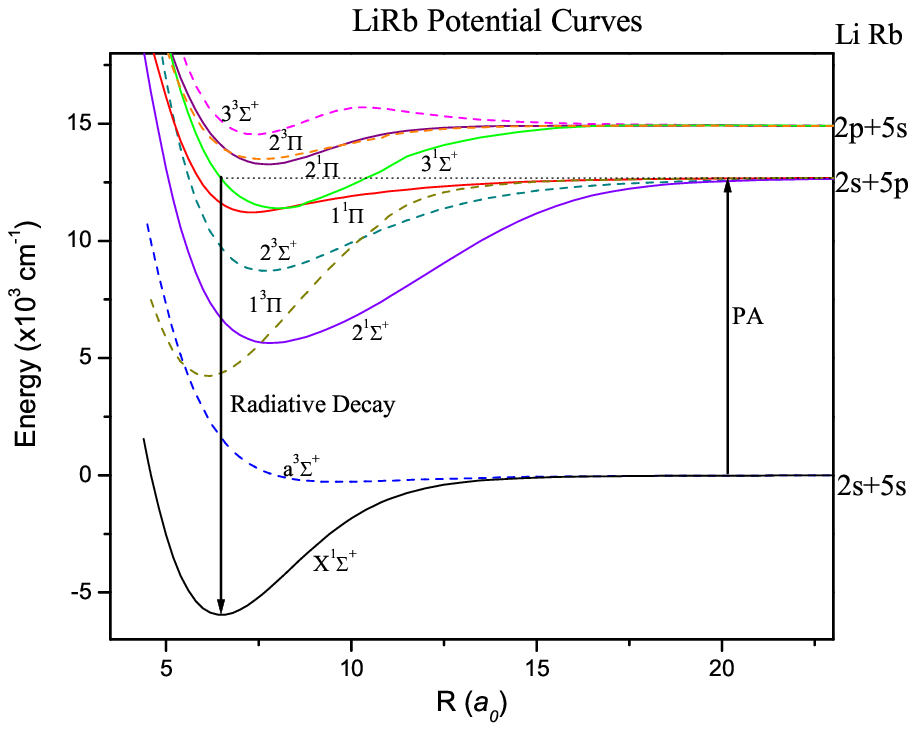}
\caption{Potential energy curves for LiRb up to the second exited atomic asymptote based on the data in Ref.\cite{Korek001}.  The vertical arrows correspond to Photoassociation (PA) and Radiative decay.  The horizontal dotted lines correspond to the energy of the lowest excited atomic asymptote.}
\label{graph:LiRb}
\end{center}
\end{figure}

Again the most promising potential energy curve is the $3 ^1\Sigma^+$ short range state which is resonantly coupled to the $1 ^1\Pi$ long range state near 8.5 $a_o$ (Fig.\ref{graph:LiRb}).  Note that here the effect of spin-orbit may be significant, slightly raising the $1 ^1\Pi$ state (which correlates to Li($2s$)+Rb(5$p_{3/2}$) with respect to the $3 ^1\Sigma^+$ state (which correlates to Li(2$p_{1/2}$)+Rb($5s$)).

The avoided crossing of the $2 ^1\Pi$ and $1 ^1\Pi$ states is no longer evident.

\subsection{  LiCs\cite{Korek00977}}

\begin{figure}[h]
\begin{center}
\includegraphics[width=4 in, height=3 in]{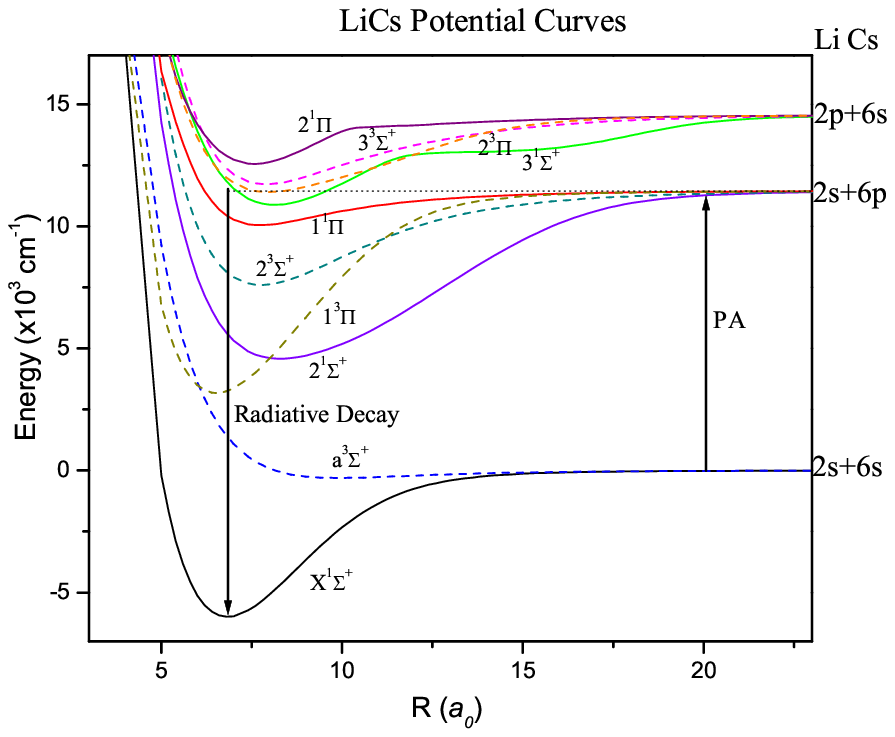}
\caption{Potential energy curves for LiCs up to the second exited atomic asymptote based on the data in Ref.\cite{Korek00977}.  The vertical arrows correspond to Photoassociation (PA) and Radiative decay.  The horizontal dotted lines correspond to the energy of the lowest excited atomic asymptote.}
\label{graph:LiCs}
\end{center}
\end{figure}

The most promising potential curve is again the $3 ^1\Sigma^+$ short range state, but now the resonant coupling is less clear, since the state does not cross the $1 ^1\Pi$ state in the absence of spin-orbit coupling (Fig.\ref{graph:LiCs}).  Given the large spin-orbit coupling in the Cs atom, it is possible that the curves will in fact cross in an {\itshape{ab initio}} calculation including spin-orbit.  We recommend that such a calculation be carried out, as it may help explain the very interesting results of Ref.\cite{Deiglmayr}.

\subsection{  NaK\cite{Magnier}}

\begin{figure}[H]
\begin{center}
\includegraphics[width=4 in,height=3 in]{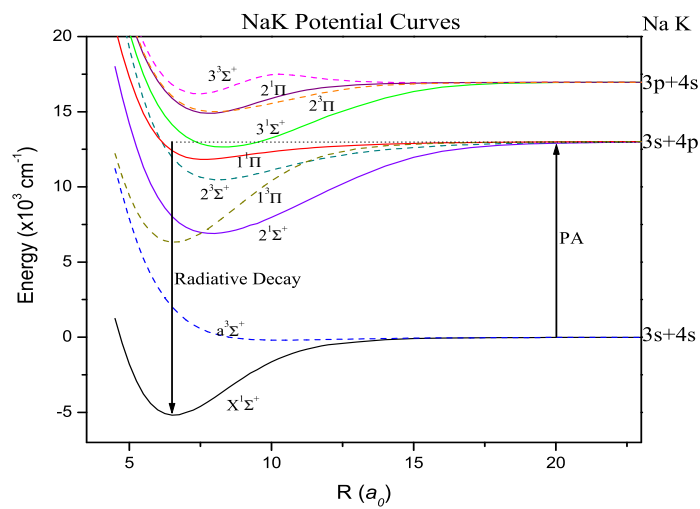}
\caption{Potential energy curves for NaK up to the second exited atomic asymptote based on the data in Ref.\cite{Magnier}.  The vertical arrows correspond to Photoassociation (PA) and Radiative decay.  The horizontal dotted lines correspond to the energy of the lowest excited atomic asymptote.}
\label{graph:NaK}
\end{center}
\end{figure}

It appears that direct PA to the $1 ^1\Pi$ state is the most promising pathway in the case of NaK (Fig.\ref{graph:NaK}).  Here detunings less than 100 cm$^{-1}$ may be required to reach X(0,0) with high probability.  It does not appear that resonant coupling will play a significant role.

\subsection{  NaRb\cite{Korek001}}

\begin{figure}[H]
\begin{center}
\includegraphics[width=4 in, height=3 in]{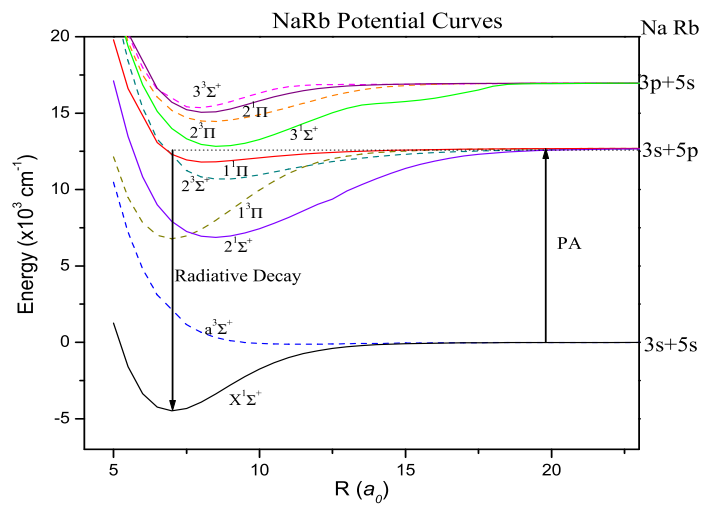}
\caption{Potential energy curves for NaRb up to the second exited atomic asymptote based on the data in Ref.\cite{Korek001}.  The vertical arrows correspond to Photoassociation (PA) and Radiative decay.  The horizontal dotted lines correspond to the energy of the lowest excited atomic asymptote.}
\label{graph:NaRb}
\end{center}
\end{figure}

Here again it appears that direct PA to the $1 ^1\Pi$ state is the most promising, even more so when the $1 ^1\Pi$ state is shifted up by the large atomic spin-orbit coupling in the Rb atom (Fig.\ref{graph:NaRb}).  It does not appear that resonant coupling will play a significant role.

\subsection{  NaCs\cite{Korek00977}}

\begin{figure}[H]
\begin{center}
\includegraphics[width=4 in, height=3 in]{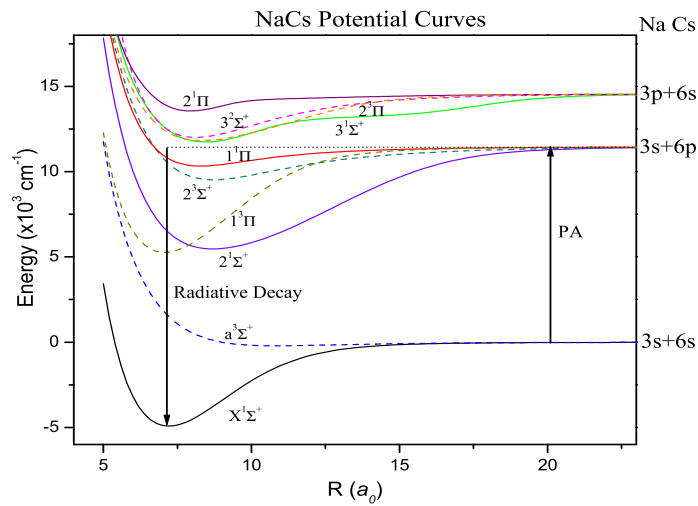}
\caption{Potential energy curves for NaCs up to the second exited atomic asymptote based on the data in Ref.\cite{Korek00977}.  The vertical arrows correspond to Photoassociation (PA) and Radiative decay.  The horizontal dotted lines correspond to the energy of the lowest excited atomic asymptote.}
\label{graph:NaCs}
\end{center}
\end{figure}

Here again it appears that direct PA to the $1 ^1\Pi$ state is the most promising pathway, even more when the $1 ^1\Pi$ state is shifted up even more than in NaRb by the large atomic spin-orbit coupling in the Cs atom (Fig.\ref{graph:NaCs}).  It does not appear that resonant coupling will play a significant role.

\subsection{  KCs\cite{Korek06}}

\begin{figure}[H]
\begin{center}
\includegraphics[width=4 in, height=3 in]{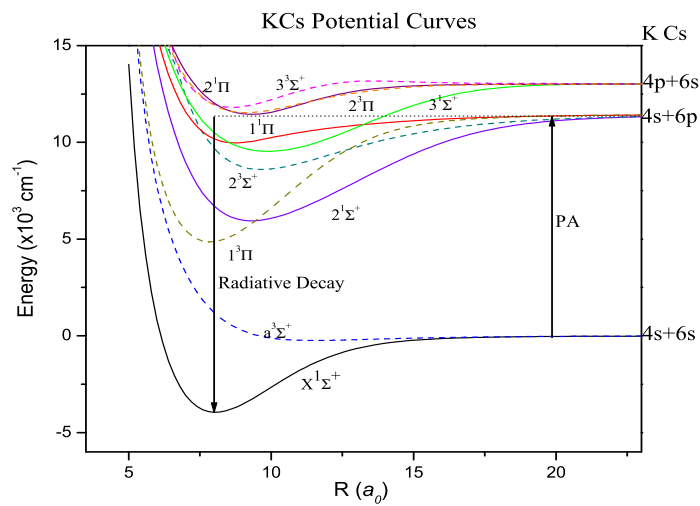}
\caption{Potential energy curves for KCs up to the second exited atomic asymptote based on the data in Ref.\cite{Korek06}.  The vertical arrows correspond to Photoassociation (PA) and Radiative decay.  The horizontal dotted lines correspond to the energy of the lowest excited atomic asymptote.}
\label{graph:KCs}
\end{center}
\end{figure}

The more promising path here appears to be the $3 ^1\Sigma^+$ $\sim$ $1 ^1\Pi$ resonant coupling as in the LiM dimers (3.1-3.4) (Fig.\ref{graph:KCs}).  However, the overlap is not as good and the reduced mass is much larger, so the extent of the X(0,0) wavefunction is much smaller, further diminishing the Franck-Condon factors. Since the spin-orbit interaction is much larger in Cs than in K, an {\itshape {ab initio}} calculation with spin-orbit included may reduce the $2 ^1\Pi$ - $1 ^1\Pi$ gap and make that resonant coupling pathway rather promising as well.  Thus a calculation involving spin-orbit coupling is desirable here.

\subsection{  RbCs\cite{Allouche}}

\begin{figure}[H]
\begin{center}
\includegraphics[width=4 in, height=3 in]{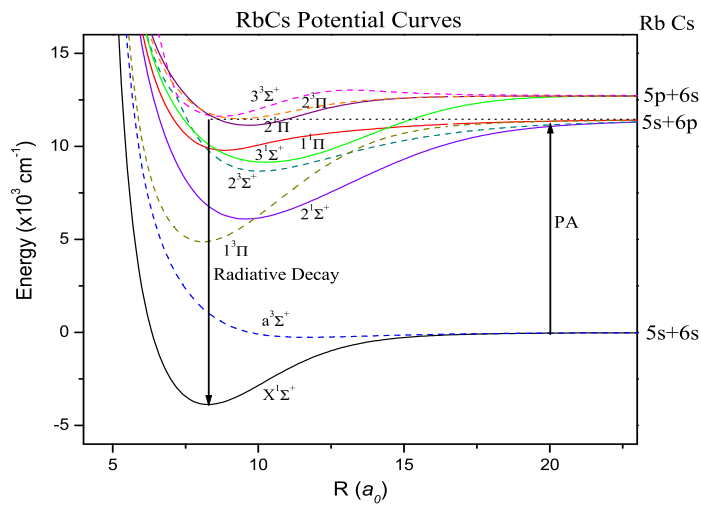}
\caption{Potential energy curves for RbCs up to the second exited atomic asymptote based on the data in Ref.\cite{Allouche}.  The vertical arrows correspond to Photoassociation (PA) and Radiative decay.  The horizontal dotted lines correspond to the energy of the lowest excited atomic asymptote.}
\label{graph:RbCs}
\end{center}
\end{figure}

As in KRb, $2 ^1\Pi$ resonant coupling with $1 ^1\Pi$ appears to be quite a promising path (Fig.\ref{graph:RbCs}).  However, a calculation involving spin-orbit coupling would also be particularly desirable in this case.

\section{Summary and Discussion}

The results of the examination of Fig.\ref{graph:LiNa}-\ref{graph:RbCs} are summarized in Table.\ref{tbl:alkali}.  Our initial example of resonant coupling of the $2 ^1\Pi$ and $1 ^1\Pi$ states in KRb also appears relevant to the case of RbCs and possibly KCs.  The resonant coupling of the $3 ^1\Sigma^+$ and $1 ^1\Pi$ states appears to be more relevant in the cases LiNa, LiK, LiRb, and possibly LiCs and KCs.  Simple direct PA appears most promising in NaK, NaRb and NaCs, without any need for resonant coupling.

\begin{table}
\caption{Promising Pathways to Formation of X(0,0) Heteronuclear Alkali
Metal Dimers}
\label{tbl:alkali}
\begin{tabular}{c c c c}\\ 
\hline
Molecule & Direct PA to 1 $^1\Pi$  &  \multicolumn{2}{c}{Resonant Coupling} \\ 
\hline
&  &  2 $^1\Pi$ $\sim$ 1 $^1\Pi$ & 3 $^1\Sigma^+$ $\sim~$1 $^1\Pi$\\
\hline
LiNa &  &   & x \\
LiK &  &   & x \\
LiRb &  &   & x \\
LiCs &  &   & x? \\
NaK  & x &   & \\
NaRb  & x &   &  \\
NaCs & x &   &  \\
KRb  &  &  x &  \\
KCs &  &  x? & x? \\
RbCs  &  &  x &  \\
\hline
\end{tabular}
\end{table}

For the direct PA cases (NaK, NaRb and NaCs), experiments as a function of the detuning of the PA laser to the $1 ^1\Pi$ state while detecting the X(0,0) level should readily identify the optimum $1 ^1\Pi$ vibrational level for forming X(0,0) molecules.

For the $2 ^1\Pi$ $\sim$ $1 ^1\Pi$ resonant coupling cases (RbCs and KCs), spectroscopy comparable to that carried out for KRb remains to be carried out.  Detecting the X(0,0) level while tuning the PA laser should again allow ready identification of the resonantly coupled levels.

For the $3 ^1\Sigma^+$ $\sim$ $1 ^1\Pi$ resonant coupling cases (LiNa, LiK, LiRb, and possibly LiCs and KCs), spectroscopy is again needed.  Detecting the X(0,0) level while detuning the PA laser should again assist in identifying the resonantly coupled levels.

In both cases of resonant coupling, there are two factors to consider:  the magnitude of the coupling of the two single channel electronic states, and the energy denominator.  The magnitudes of the short range couplings of two singlet states discussed in Section III probably do not vary greatly, although the homogeneous perturbations ($\Delta {\Lambda}$ = 0) are often somewhat larger than the heterogeneous ones ($\Delta {\Lambda}$ = $\pm$1).  However, we have thus far considered electronic states to be eigenfunctions of total electronic spin S.  However, the long range states formed by PA are generally not eigenfunctions of S, but rather of J$_z$, with eigenvalue $\Omega$, the projection of the total electronic angular momentum along the internuclear axis z.  Thus for a particular lowest excited asymptote of a heteronuclear alkali dimer (say {\itshape{n}}$s$ + {\itshape{n}}$'$ $p_{1/2}$), there are three states ($0^+$, $0^-$, and 1) and for the second lowest excited asymptote ({\itshape{n}}$s$ + {\itshape{n}}$'$ $p_{3/2}$), there are five states ($0^+$, $0^-$, 1, 1, and 2).  While the $0^-$ and 2 states are pure triplets, the other states are each singlet/triplet mixtures.  Moreover, different states of the same $\Omega$ value, e. g. the three $\Omega$=1 states, can also resonantly couple, as can states with ($\Delta {\Omega}$ = $\pm$ 1).  Thus the $1 ^1\Pi$ state we speak of being formed by PA (actually a 1 state correlating with the second lowest excited asymptote) may not be a pure singlet state for resonant coupling.  Nevertheless, in the region inside 15 $a_o$, it is well described as a singlet state.  The states to which the resonant coupling occurs ($2 ^1\Pi$ and $3 ^1\Sigma^+$) are in fact well described as singlets as long as they are well below their dissociation limits (to the fourth and third lowest excited asymptotes, respectively), which they always are in the PA region below the lowest excited asymptote.

The other factor, the energy denominator, can greatly increase the strength of resonant coupling when very small.  The odds of very small energy denominators scale with the density of vibrational levels, i. e. with the square root of the reduced mass.  Thus strong resonant coupling corresponding to very small energy denominators is most favored for RbCs and least favored for LiNa; however, this is just a factor of 3.3 in square root of the reduced mass. In $^{39}$K $^{85}$Rb \cite{Wang}, for example, in addition to the (60,17) pair of resonantly coupled levels shown in Fig.\ref{graph:spectrum}, there is the (64,18) pair $\sim$2 cm$^{-1}$ above the K($4s$) Rb($5p_{1/2}$) asymptote and the (57,16) pair $\sim$87 cm$^{-1}$ below that asymptote.  Thus the number of pairs of resonantly-coupled vibrational levels will probably remain in the single digits for all ten systems.

It might also be noted that other useful comparisons among the heteronuclear alkali dimers have previously been published:  a comparison of the strength and signs of the long range potentials and the probability of PA\cite{Wang98},  a comparison of the theoretical and experimental dipole moments\cite{Aymar}, a comparison of photoassociation and molecule formation probabilities  \cite{Azizi} and a comparison of stimulated Raman schemes for forming X(0,0) molecules via the resonantly coupled $2 ^1\Sigma^+$ $\sim$ $1 ^3\Pi$ states starting in a magnetoassociated state of predominantly triplet character\cite{Stwalley}.

\section{Acknowledgement}

Helpful discussions with Hyewon Kim Pechkis, Ed Eyler, and Phil Gould are gratefully acknowledged, as is support from the National Science Foundation. The original data used in Fig.\ref{graph:spectrum} were obtained by Dajun Wang. We also thank Olivier Dulieu for providing his calculations for the potential energy curves of LiK in advance of publication, and Marko Gacesa, Philippe Pellegrini, and Robin Cote for discussion of FOPA and its relation to the resonant coupling of excited states discussed herein.We also thank Ed Eyler for his careful reading of the manuscript and helpful suggestions.

\end{document}